\documentclass[aps,prl,twocolumn,groupedaddress]{revtex4}
\usepackage{graphicx}
\usepackage{dcolumn}

\begin{document}
\title{Combined photonic-plasmonic modes inside photonic crystal cavities}
\author{Authors: Abelali Mrabti, Sa\"id El-Jallal, Ga\"etan L\'ev\^eque,\\ Abdelatif Akjouj, Yan Pennec, and Bahram Djafari-Rouhani}
\affiliation{Institut d'\'Electronique, de Micro\'electronique et de Nanotechnologie, 59652 Villeneuve d'Ascq Cedex, France}
\email{Email: gaetan.leveque@univ-lille1.fr}

\begin{abstract}
In this article, we present a numerical study of the optical properties of a metal nanowire interacting with a localized cavity in a two dimensional photonic crystal. The nature of the modes, their wavelength and width are investigated as a function of the particle radius. It is shown in particular that a particle with diameter about the lattice constant presents very narrow resonances corresponding to hybrid photonic-plasmonic modes, where increased lifetime is attributed to the decrease of the radiative losses by interaction with the photonic crystal. These results open interesting applications in areas where narrow plasmonic resonances are required, as in LSP-resonance based biosensing.
\end{abstract}

\maketitle

\section{Introduction}

Localized surface plasmons (LSPs) are optical modes supported by metal nanoparticles, which result from the resonant coupling between the oscillating surface charge created along the particle's surface by an incident electromagnetic field, and the light wave that it scatters \cite{Maier2007}. For a specific wavelength, depending on the particle shape, volume, material, as well as the optical characteristics of its environment, the extinction cross-section increases, and the intensity of light can be enhanced by several orders of magnitude in a length range around the particle well below the wavelength of the incident light \cite{Martin2001}. For all of these properties, LSPs are investigated due to both fundamental aspects and applications into numerous domains like biosensing \cite{Mock2003,saison2012plasmonic,Maurer2013}, surface enhanced Raman spectroscopy \cite{Wen2014}, photo-thermal therapy \cite{Jaque2014}, or metamaterials and optical cloaking \cite{Liu2011}. In number of applications, its is required to have a resonance as narrow as possible, for example in biosensing where the principle is to detect a small shift in the resonance wavelength when a target molecule is adsorbed on a metal nanoparticle surface: the resonance width determines the resolution of the sensor, and is often the main limitation in LSP-resonance-based sensing \cite{Mayer2011}. As the width of the plasmon resonance is determined by ohmic
(which cannot be suppressed) and radiative losses, the narrowest resonances are obtained with the so-called "dark" plasmons modes: the distribution of electric charges on the particle surface is such as the particle does not possess any net dipole, and the radiative waves are cancelled out \cite{Schmidt2012,Verellen2011}. Narrow plasmon resonances can be obtained as well by tailoring the environment of the particle, for example by making it interact with photonic crystals (PCs), as these structures present forbidden bands for radiative waves, which can be tuned by playing with their geometrical parameters. Recently, Solano and co-workers have shown \cite{Sanchez-Sobrado2011,Solano2012} that the absorbance of coated gold nanospheres and nanorods could be tailored when placed in a defect layer of a multilayer planar resonator: enhancement of the absorption takes place within the band-gap at wavelengths close to the one of the cavity defect, and narrowing of the resonance results from the filtering of light by the cavity mode. Following that work, Wang et al \cite{Wang2013} have shown theoretically that the same system could show greatly enhanced photo-induced change of the optical transmittance, based on the modification of the dielectric function of the gold nanoparticles on a subpicosecond time scale with an ultrashort light pulse. 
In that perspective, increasing the confinement of the localized plasmon mode of a nanoparticle using two-dimensional photonic crystal is of great interest for novel photonic devices. Hence, we propose in this article a detailed study of a 2D photonic crystal interacting with an arbitrary size 2D metal nanoparticle.
\begin{figure}
\centering
\includegraphics[width=6.5cm]{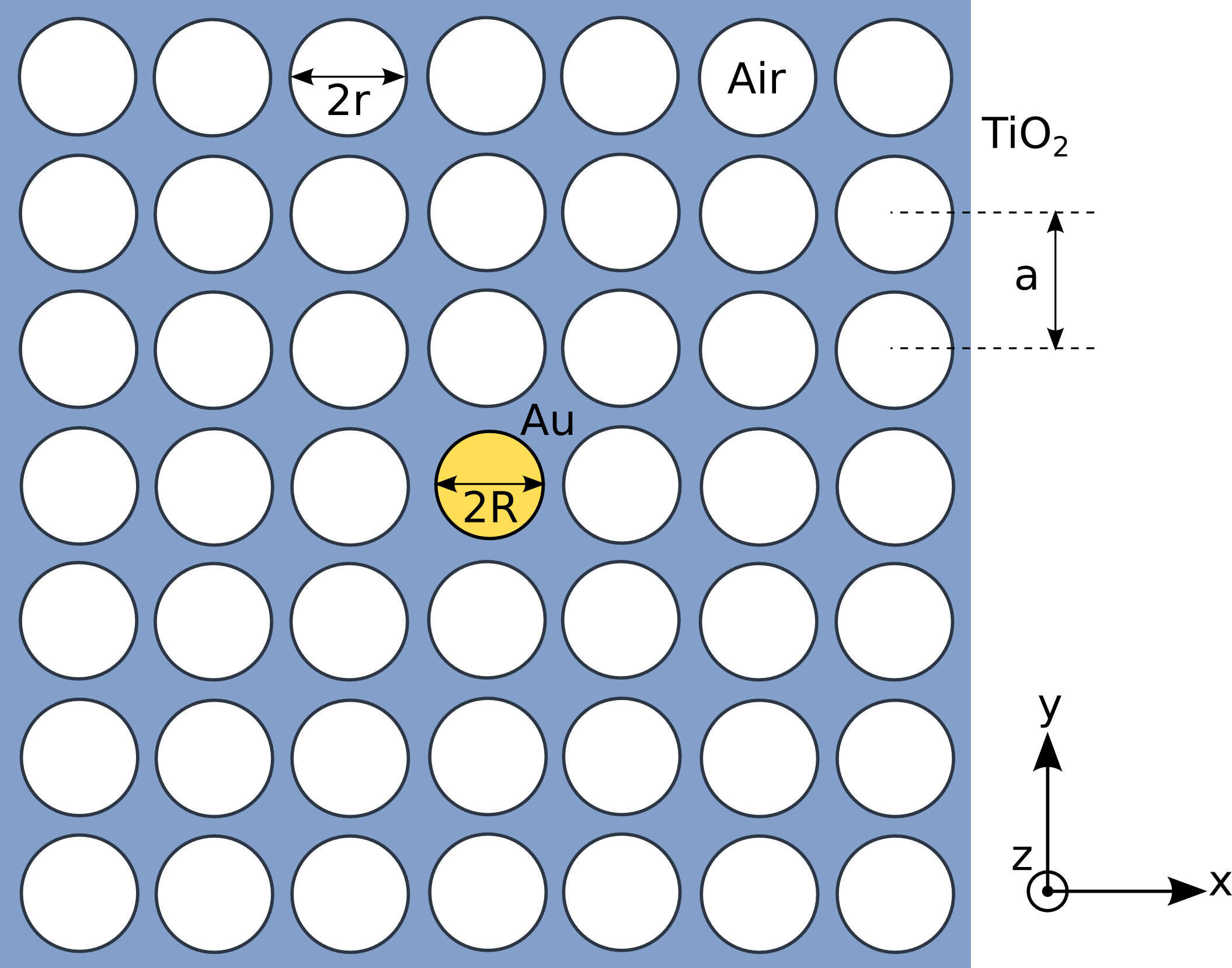}
\caption{The investigated structure is a square photonic crystal with lattice parameter $a$, made of air-holes with radius $r$ inside a TiO$_2$ matrix, where one of the hole is replaced by a metal nanowire of varying radius $R$.}
\label{fig:fig1}
\end{figure}

More precisely, we investigate in this article the optical properties of a two-dimensional metal nanowire (NW) placed inside a localized cavity of a photonic crystal (Figure \ref{fig:fig1}). We aim at characterizing the optical modes of that system, by their nature, wavelength and resonance width, together with their electromagnetic field distribution. The geometrical parameters of the photonic crystal have been chosen in order to have the localized plasmon wavelengths of the particle within the band-gap of the PC. We expect that the radiative contribution to damping rate will be strongly modified as the radiative waves emitted by the nanowire at the resonance are prevented from leaking inside the PC.
The PC is made of a square grating of 2D circular air holes of radius $r$ and lattice parameter $a$, embedded inside a titanium dioxyde (TiO$_2$) matrix. This material has been chosen because of its transparency in the visible range of wavelengths, where its refractive index has been taken as constant and equal to 2.9 in the following. That high value allows ensuring a large index contrast between the material and the air holes, which allows obtaining large and deep band-gaps, that could not be reached with lower index materials like SiO$_2$. The metal 2D nanowire (NW) is a cylinder of radius $R$ made of gold, and is placed at the center of a localized cavity obtained by filling an air hole with TiO$_2$.
The moderate refractive index of TiO$_2$ allows keeping the surface plasmon wavelengths of the nanowires in the visible range. The data for the dielectric constant of gold are extracted from Johnson and Christy \cite{Johnson1972}, and the calculations were performed using finite element methods (Comsol Multiphysics, Radio-Frequency module). In the first part, we present the properties of the gold nanowire embedded in an homogeneous TiO$_2$ medium. In parallel, we show the band-gap tuning properties of the photonic crystal as a function of the grating filling factor, together with the calculation of the photonic modes sustained by the localized cavity. In the second part, we present how both the photonic modes of the PC and the particle's plasmonic modes are affected when the metal particle is placed inside the cavity. The results are discussed as a function of the radius of the gold NW, first when it is much lower than the grating's period, and second when both the radius and the period have similar values.
\section{Optical properties of the metal nanowire and the uncoupled PC}
\subsection{Au nanowire inside an homogeneous and infinite TiO$_2$ medium}
\begin{figure}[h]
\centering
\includegraphics[width=5.5cm]{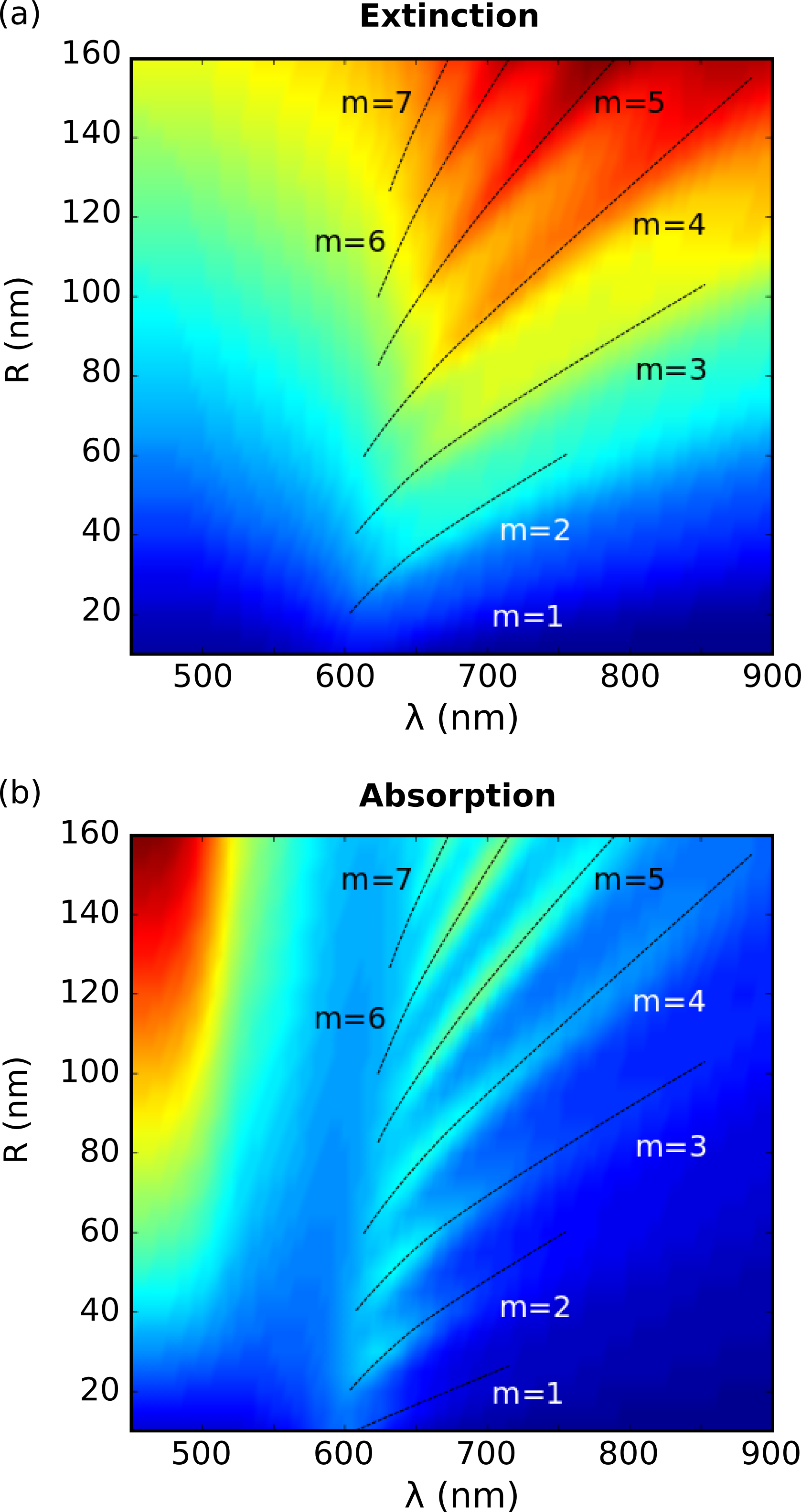}
\caption{Extinction (a) and absorption (b) plotted as a function of the wavelength and the nanowire's radius (arbitrary units, maximum in red, minimum in blue). The labels $m$ indicate the order of the different localized plasmon mode.}
\label{fig:fig1b}
\end{figure}

\begin{figure}
\centering
\includegraphics[width=6.5cm]{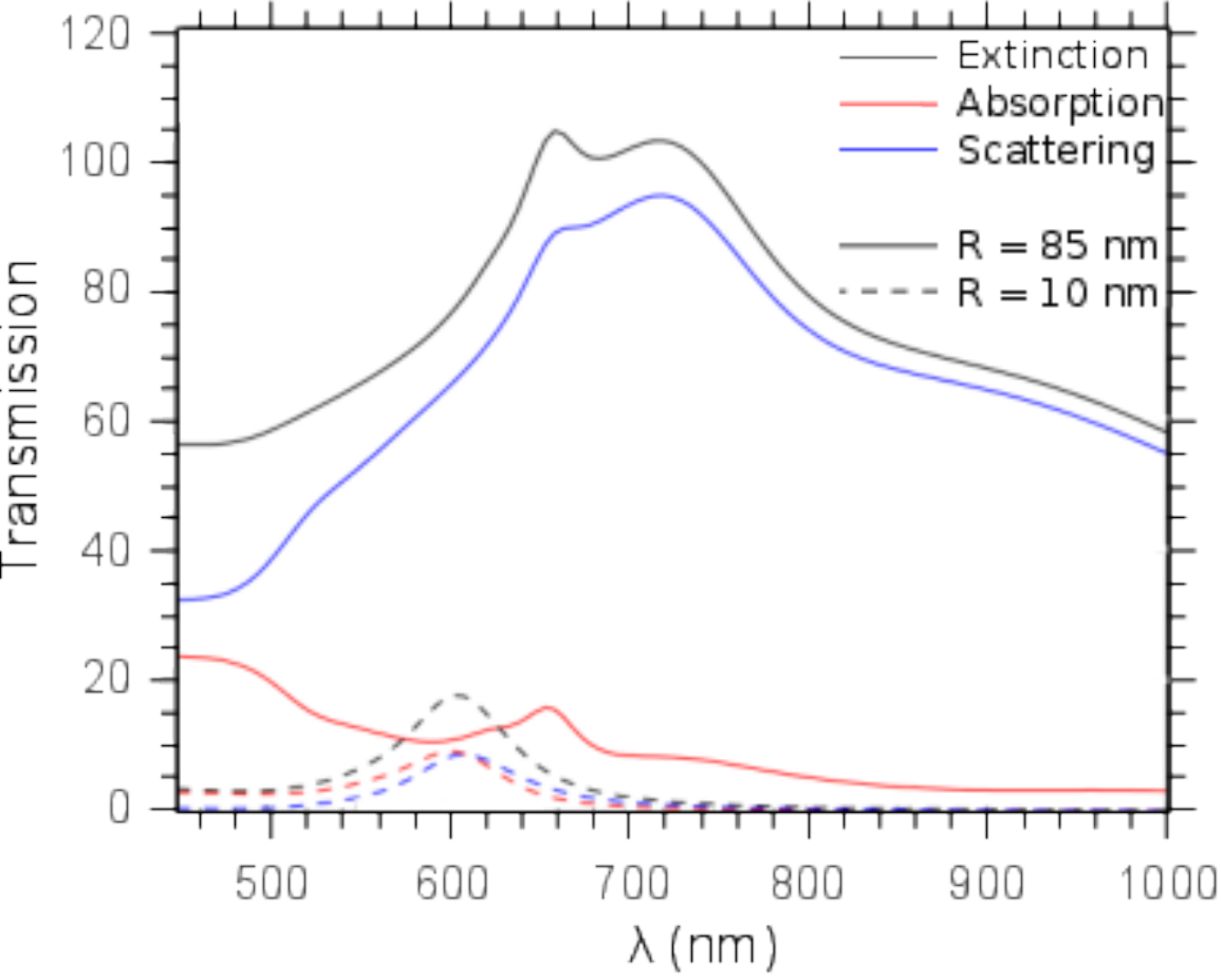}
\caption{Comparison between the extinction (black), absorption (red) and scattering (blue) cross-sections for a particle of radius $R=10$ nm (dashed line) and a particle of radius $R=85$ nm (solid line).}
\label{fig:fig1c}
\end{figure}

Figure \ref{fig:fig1b} shows the extinction and absorption spectra of the gold nanowire in the homogeneous TiO$_2$ matrix as a function of its radius $R$. The particle have several high-order plasmon resonances which have their largest extinction signature between 600 and 700 nm. The resonances are labelled by their order $m$, $m=1$ corresponding to the dipolar mode, $m=2$ to the quadrupolar mode and so on. As the refractive index of the embedding medium is quite high compared to more usual materials in plasmonics, like silica or water, the LSP modes supported by the particle can have a large order $m$, up to $m=7$ for the 150 nm radius particle. The quadrupolar mode appears in the visible range of wavelengths for a diameter of about 40 nm, for which the dipolar resonance is red shifted and barely noticeable. The absorption map (Fig.~\ref{fig:fig1b}(b)) is quite different from the extinction map (Fig.~\ref{fig:fig1b}(a)), due to from the fact that the contribution of the scattering to the extinction increases with the particle size. As can be seen on Fig.~\ref{fig:fig1c}, the absorption is comparable to the scattering for a small particle of radius about 10 nm, but, as the particle is rapidly large compared to the wavelength of the light in the medium (far from the electrostatic approximation), the scattering largely dominates when the particle diameter increases, as it can be verified for $R=85$ nm. This is the reason why in Fig.~\ref{fig:fig1b} the extinction is high for larger radii, while the absorption appears low.
\subsection{Photonic crystal}
\begin{figure}
\centering
\includegraphics[width=5.5cm]{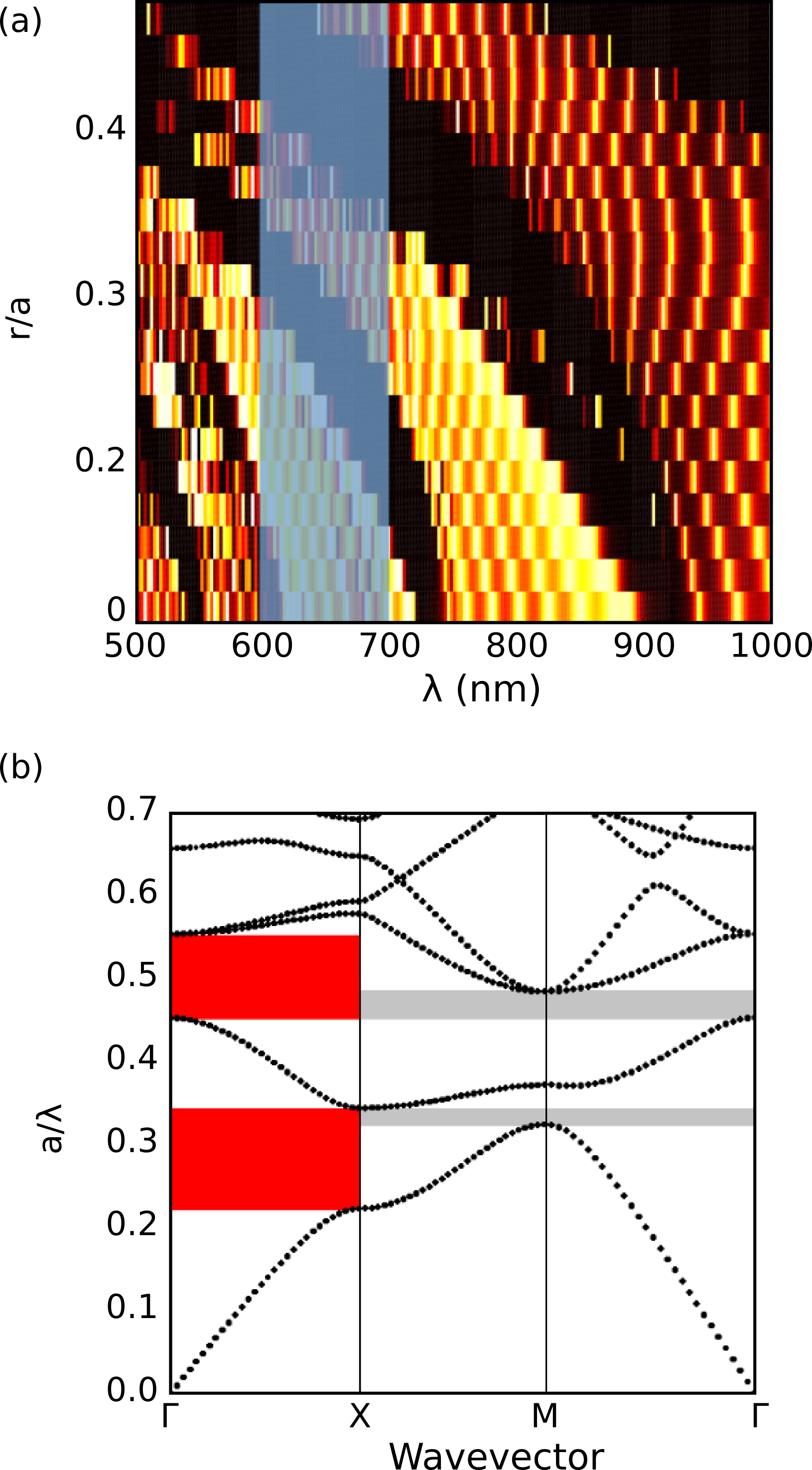}
\caption{(a) Transmission of a TM-planewave in normal incidence on a PC made of seven lines of air holes with radius $r$ and period $a=320$ nm, as a function of $r/a$. Maximum transmission is white, minimum is black; (b) Band diagram of the PC along the high symmetry axes $\Gamma X$, $XM$ and $M \Gamma$ of the first Brillouin zone, with period $a=320$ nm and $r/a=0.42$. The gray areas represent the absolute band gap whereas the red areas represent the band gap in the $\Gamma X$ direction.}
\label{fig:fig2}
\end{figure}

The geometrical parameters of the photonic crystal have been chosen in order to tune its absolute band-gap to the 600 nm / 700 nm wavelength domain, where the nanowire's LSP modes are the most efficiently excited. As the refractive index is fixed, the only two degrees of freedom are the period $a$ of the grating and the radius $r$ of the air hole, which both allow to change the central wavelength, the width and the depth of the band gap. We have fixed the period to $a=320$ nm, while figure \ref{fig:fig2}(a) shows the evolution with $r/a$ of the transmission spectra of a planewave in normal incidence onto one of the (10) direction of the grating, limited to 7 lines of holes. The polarization of the incident wave is TM, the incident magnetic field being along the invariance direction. In this transmission calculations and the following, periodic boundary conditions have been applied in the $x$-direction, while perfectly matched layers have been applied in the incidence and emergence media, perpendicular to the incident planewave direction. The transmission spectrum has been computed by integrating the Poynting vector flow through a line parallel to the PC in the emergence medium. 
It appears that the structure supports several band gaps, and that for $R/a\approx0.42$ a large band gap overlaps with the target wavelength domain (blue area). In order to check the consistency of the transmission calculations, we have compared these results with a full modal dispersion calculation, as shown on Fig.~\ref{fig:fig2}(b), for $a=320$ nm and $r=0.42 a$. We can see that a complete band gap exists in the reduced frequency domain $a/\lambda\in [0.4524;0.4841]$ ($\lambda\in [661\mbox{ nm};$ $707\mbox{ nm}]$ with $a=320$ nm), whereas a partial band gap in the $\Gamma X$ direction exists in the interval $a/\lambda\in [0.452;0.554]$ ($\lambda\in [578\mbox{ nm};707\mbox{ nm}]$ with $a=320$ nm).
\begin{figure}
\centering
\includegraphics[width=8.5cm]{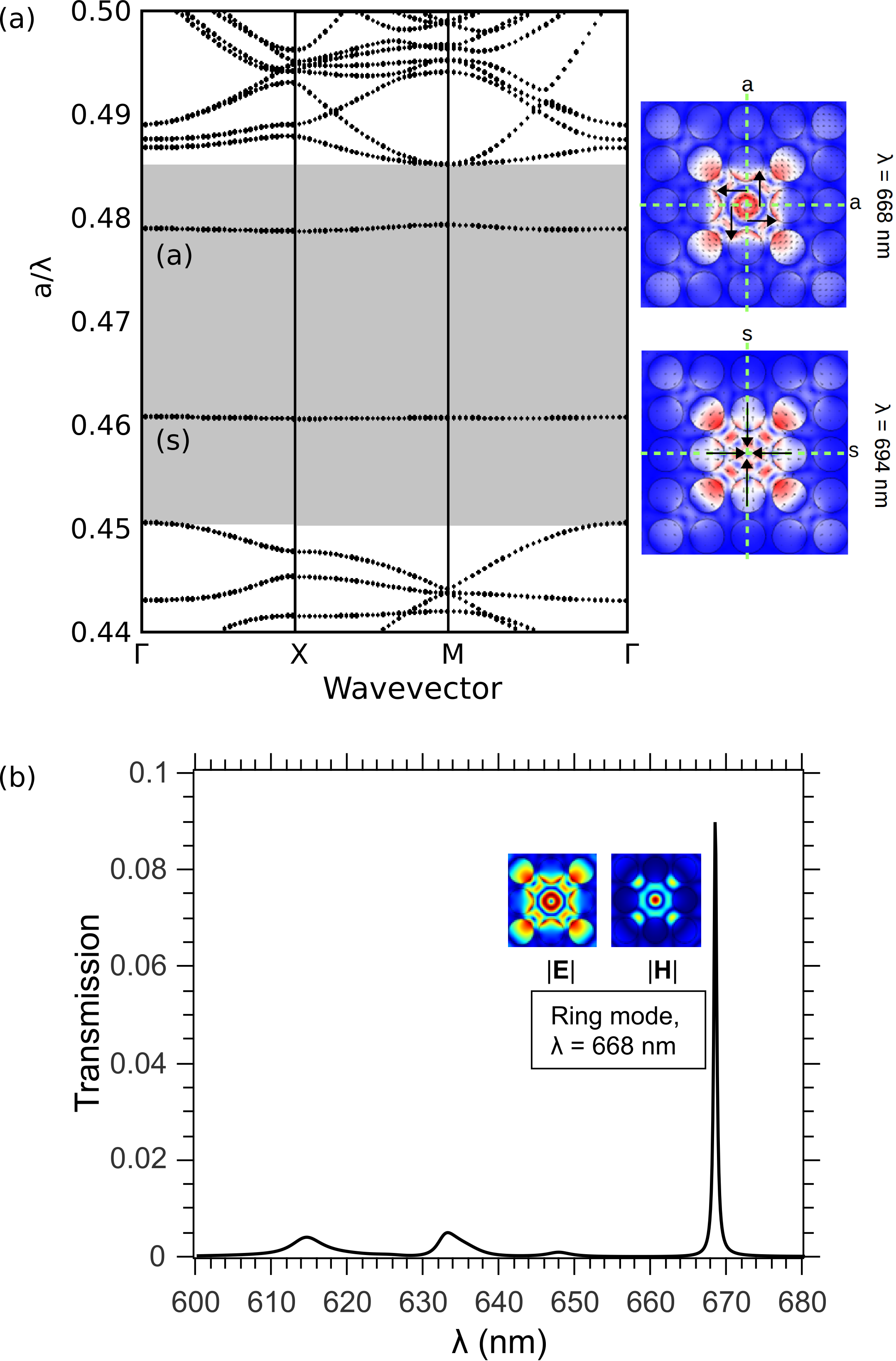}
\caption{(a) Band diagram of the 7x7 supercell of the same PC with a localized cavity. The wavelength and electric field distributions have been indicated for a = 320 nm; (b) Transmission of a TM-planewave in normal incidence on the same PC, limited to 7 lines along the direction of the incident wave.}
\label{fig:fig3}
\end{figure}

A cavity is then created by filling one of the air holes with TiO$_2$. On Figure \ref{fig:fig3}(a), the dispersion calculation on a 7x7 supercell shows that two localized photonic modes, respectively (a) and (s), are found inside the complete band-gap for reduced frequencies $a/\lambda$ respectively equal to 0.48 ($\lambda=668$ nm)  and 0.46 ($\lambda=694$ nm). The distribution of the electric field amplitude is shown on the colormaps, Fig.~\ref{fig:fig3}(a), right. The lowest wavelength mode (a) has an annular shape, its field lines form closed loops around the center of the cavity where the intensity reaches a minimum. We can notice that the mode is antisymmetric with respect to the two principal axes of the photonic crystal cavity. The highest wavelength mode (s) has a very different distribution of electric field, and shows several local maxima inside the cavity. It is symmetric with respect to the main axes of the photonic crystal cavity and reaches as well a minimum amplitude at the center of the cavity. Figure \ref{fig:fig3}(b) shows the transmission spectra of a TM-planewave normally incident on the (10) direction of the grating, still limited to 7 rows. The wavelength and electric field distribution of the main resonance is consistent with the ring mode (a) found with the dispersion calculation. The mode at 694 nm is not excited, because, as the incident light is normally incident onto the photonic crystal, it is antisymmetric with respect to the vertical direction, and cannot couple to the 694-nm-symmetric mode. However, the same calculation performed with a small incidence angle breaks the symmetry of the incident field and allows to recover the corresponding transmission peak (not shown) at the wavelength predicted by the dispersion calculation. 

\section{Coupled systems}
In this section, we investigate the optical behaviour of the gold nanowire placed at the center of the localized PC cavity. The absorption spectra have been calculated as a function of the particle radius, by integration of the power losses on the cross section of the nanowire, and normalization to the incident energy flow on the 7x7 supercell. Hence, reflection ($R$), transmission ($T$) and absorption ($A$) obeys the energy conservation $R+T+A=1$. Two very different behaviours have been obtained, according to the radius of the particle compared to the grating's period.
\begin{figure}[h]
\centering
    \includegraphics[width=8.5cm]{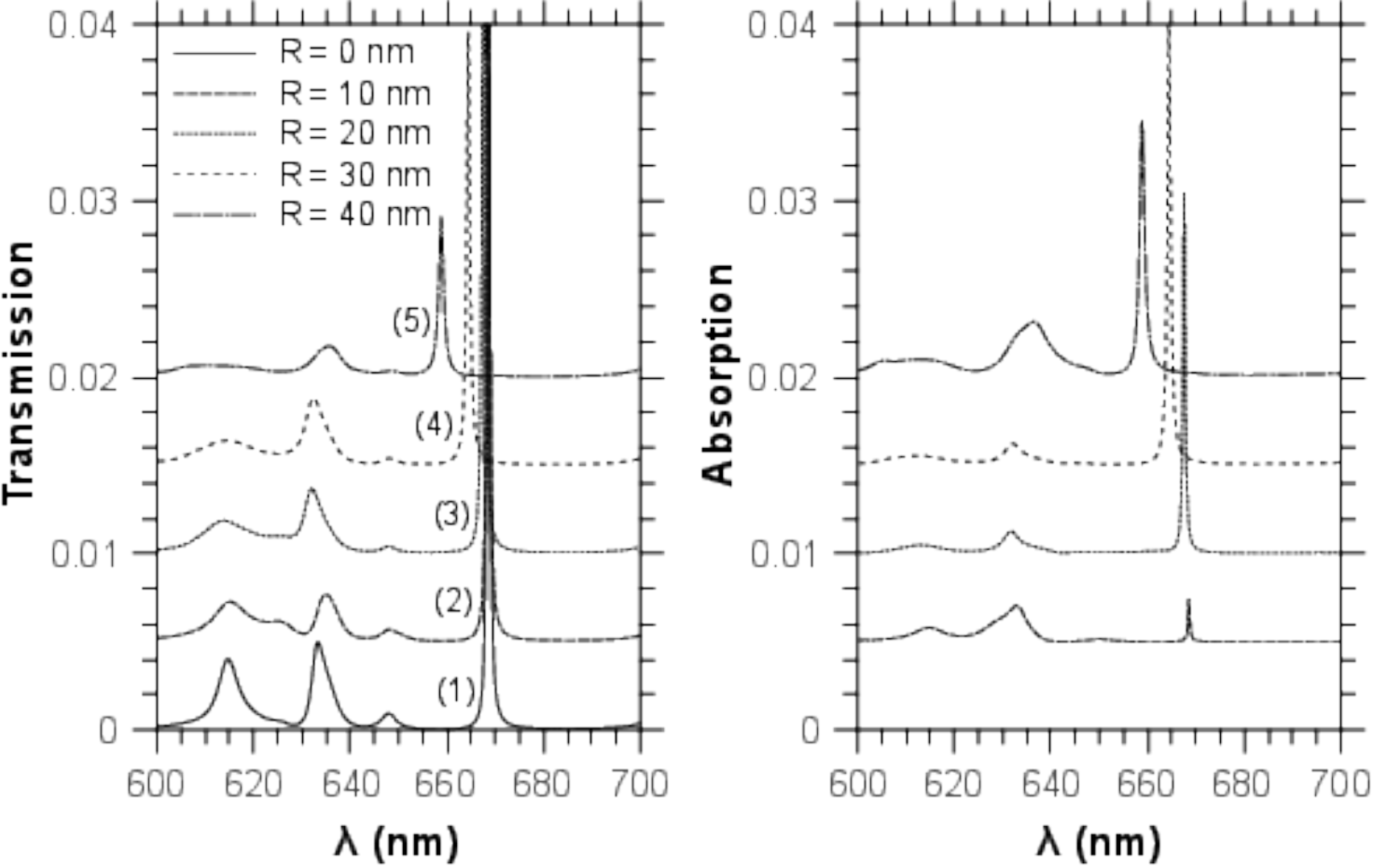}
    \caption{Transmission and absorption spectra of the system composed of the gold NW of radius $R$ placed at the center of the cavity, for wavelengths close to the cavity ring-mode position. For readibility purpose, spectra have been shifted by 0.005 per radius increment. Maxima of transmission(absorption) are: 8.5(0.24)\% (R = 10 nm), 6.0(2.0) \% (R = 20 nm), 2.5(2.6) \% (R = 30 nm), and 0.9(1.5) \% (R = 40 nm).}
    \label{fig:fig5}
\end{figure}

\subsection{Case where $R<<a$}
We consider first the case where the particle has a radius $R$ small compared to the lattice parameter $a$. Figure \ref{fig:fig5} shows the evolution of the transmission peak of the ring mode (a), from radius 10 to 50 nm. The wavelength of the mode is blue-shifted when $R$ increases, while its transmission rapidly decreases. For comparison, the absorption spectra have been added, which shows that both absorption and transmission reach comparable low values, and that most of the energy is then reflected. Table \ref{tab:tab1} shows the different resonant wavelengths, together with the electric and magnetic field intensity distribution for $R \le 40$ nm. It appears that the particular ring structure of the unperturbed PC cavity modes persists for values of $R$ up to few nanometers. Indeed, as the intensity of the PC-cavity ring mode is almost zero at center, its shape and wavelength is unchanged in the limit of very small particle ($R=10$ nm). However, when $R$ increases, we can intuitively understand the decrease of the mode wavelength by the fact that the light is pushed away from the area occupied by the metal particle: this leads to a decrease of the area accessible to the photonic mode, and then a shortening of the resonance wavelength. 
\begin{center}
\begin{table}
\centering
\begin{tabular}{>{\centering}m{1cm}|>{\centering}m{1cm}>{\centering}m{2.cm} >{\centering\arraybackslash}m{0.5cm}}
 & {\bf $R$(nm)} & $|\mbox{\bf E}|\quad$ $|\mbox{\bf H}|$ & {\bf $\lambda$(nm)} \\
 	(1) & 0     & \includegraphics[width=1.5cm]{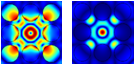} & 668.6            \\
 	(2) & 10     & \includegraphics[width=1.5cm]{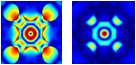} & 668.5            \\
 	(3) & 20     & \includegraphics[width=1.5cm]{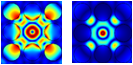} & 667.6            \\
 	(4) & 30     & \includegraphics[width=1.5cm]{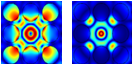} & 664.5            \\
 	(5) & 40     & \includegraphics[width=1.5cm]{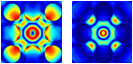} & 658.9
  \end{tabular}
 \caption{Summary of the wavelengths and field distributions for the different peaks labelled (1) to (5) on Fig.~\ref{fig:fig5}.}\label{tab:tab1}
  \end{table}
\end{center}
Moreover, if the shape of the mode is unchanged for small radius, it is not the case for a particle radius larger than 50 nm, where the light ring is much closer to the adjacent air holes. This breaks the revolution symmetry of the ring mode, and for $R>50$ nm the ring mode does not exists anymore. Finally, let us emphasize that this mode, when the particle radius is much smaller than the lattice constant, is purely photonics, without plasmonic characteristics, as the electric field is, in every point of the metal interface, parallel to it, and then cannot create the surface charge necessary to excite surface plasmon waves.
\subsection{Case where $2R$ is comparable to $a$}
The opposite regime is when the particle diameter is comparable to the lattice parameter. In that case, the light is squeezed in the very narrow space in between the gold interface and the air holes, and strong coupling occurs between the gold nanowire and the photonic crystal.
\begin{figure}[h]
    \includegraphics[width=8.5cm]{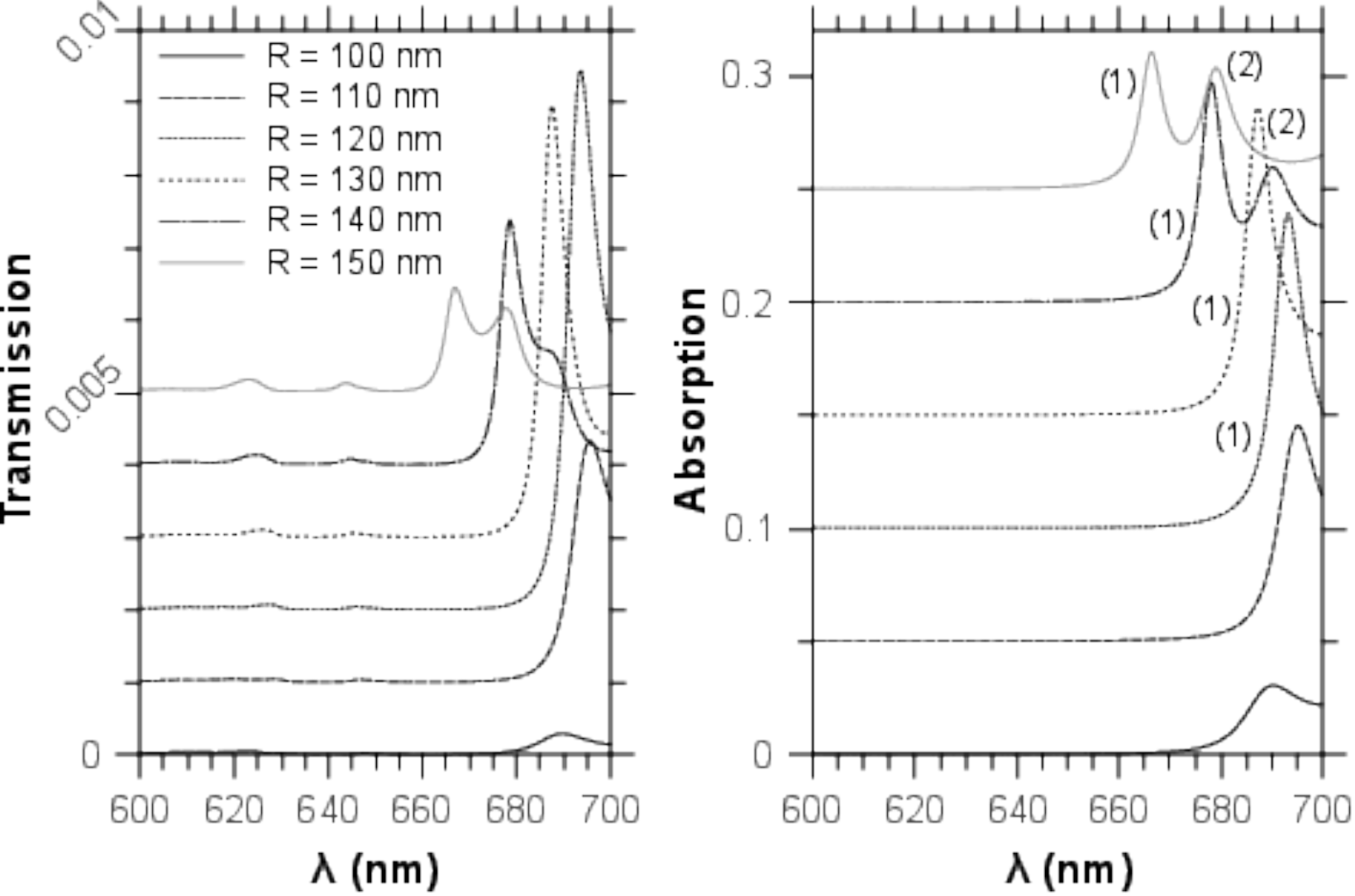}
    \caption{Transmission and absorption spectra of the coupled system with the NW's  radius $R$ comparable to the grating's period $a$. For readability purpose, spectra have been shifted by 0.001 per radius increment for transmission and 0.5 per radius increment for absorption. Maxima of transmission(absorption) are: 0.03(3.0)\% (R = 100 nm), 0.33(9.5)\% (R = 110 nm), 0.75(13.9) \% (R = 120 nm), 0.60(13.6) \% (R = 130 nm), 0.34(9.7) \% (R = 140 nm) and 0.15(6.1) \% (R = 150 nm).}\label{fig:fig6}
\end{figure}

Figure \ref{fig:fig6} shows the evolution of the transmission and absorption spectra for particle radius varying from 100 nm to 150 nm. Maximum values, indicated in the caption, show that the incident energy is mostly reflected, but a significant part is absorbed (up to 14\% for R = 120 nm) whereas less than one percent is transmitted. This is a very different behaviour than for small radius, where both maximal absorption and transmission are of the same order of magnitude. For $R$ between 100 and 130 nm, only one mode is excited between 600 and 700 nm, labelled (1). It is red-shifted with $R$ until $R=110$ nm, while for larger radius it blue shifts. Similarly, a second peak, labelled (2), appears for $R=140$ nm at a wavelength of about 660 nm, and is blue-shifted with $R$. This blue-shift of these two modes with large values of $R$ might be surprising as the localized plasmon modes of a single nanowire usually shift to the red with diameter (see Fig.~\ref{fig:fig1}). This is actually what is observed when $R<110$ nm. when $R$ becomes larger, the reversed evolution of the wavelength can be explained by the decrease of the effective refractive index surrounding the particle: for wider NW, the air holes are closer to the wire's edges and the portion of low refractive index (air holes) in the area occupied by the optical mode is larger. In that case, the blue shift due to the lowering refractive index is larger than the red shift due to the increasing diameter.
\begin{center}
\begin{table}
\centering
\begin{tabular}{>{\centering}m{0.5cm}|>{\arraybackslash}m{6cm}}
 &$\qquad|\mbox{\bf E}|\qquad\quad$\, $|\mbox{\bf H}|\quad\quad$ Surf. charges \\
 	(1) & \includegraphics[width=5cm]{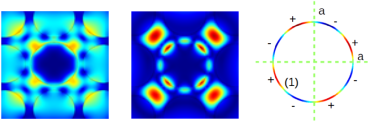} \\
 	(2) & \includegraphics[width=5cm]{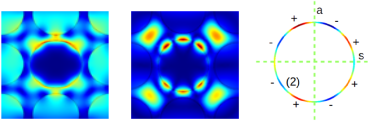}
  \end{tabular}
 \caption{Structure of the compounds modes (1) and (2) for $R=150$ nm.}\label{tab:tab2}
  \end{table}
\end{center}

The distribution of the electric and magnetic field amplitude, together with surface charges (plotted at the time of maximal amplitude) are displayed on Table \ref{tab:tab2} for the modes (1) and (2) and $R=150$ nm. In both cases, the distribution of the electric field looks complex and difficult to interpret. However, the magnetic field structure is much simpler and shows clearly a similarity between the mode (1) and the octupolar mode ($m=4$) of the single nanowire in the homogeneous TiO$_2$ matrix: the main qualitative difference is at the proximity of the air holes where one intensity maxima out of two around the particle is lower than its nearest neighbors. The similarity with the octupolar mode is confirmed by the repartition of the surface charges which clearly demonstrates that this mode is antisymmetric with respect to the two principal axis of the photonic crystal.

The mode (2) has a very different structure as indicated by the magnetic field distribution, as the maxima around the nanowire surface are evenly distributed, except on the axis perpendicular to the excitation direction where the light intensity is zero. This can be understood by the surface charges distribution which is antisymmetric compared to the vertical main axis whereas it is symmetric compared to the horizontal axis. This results in a zero of field along the nanowire's horizontal diameter. This last mode is clearly different from the single nanowire modes and as a direct consequence results intrinsically from the coupling of the localized plasmon of the particle with the photonic crystal. As both modes (1) and (2) result from the strong coupling of the NW and the PC, we will call them PC-plasmon modes in the following. 
\begin{figure}[h]
    \includegraphics[width=8.5cm]{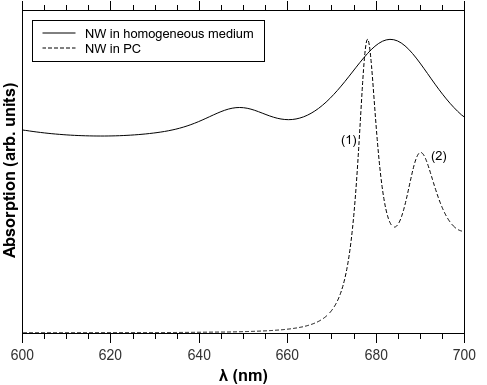}
    \caption{Comparison between the absorption spectra for the single nanowire in infinite TiO$_2$ medium (a) and the nanowire of radius 140 nm in the PC cavity.}
    \label{fig:fig7}
\end{figure}

We compare on Fig.~\ref{fig:fig7} the modes (1) and (2) for the $R=140$-nm particle (a) with the modes of the isolated nanowire with the same radius in the homogeneous matrix (b). We can clearly see that the PC-plasmon modes are much narrower. More precisely, mode (1) and (2) have a similar full width at half maximum (FWHM) of about 6 nm, which corresponds to a quality factor of 113 for mode (1), whereas the $m=6$ mode at 680 nm for the isolated nanowire has a FWHM of 36 nm with a quality factor of 19. This reduction of the mode width by a factor of about 6 is attributed to the lowering of the radiative contribution of the electromagnetic field scattered by the nanowire at the resonance, which is forbidden to propagate into the photonic crystal. This phenomenon should be responsible for the decreases of the radiative losses, making the plasmon mode width limited principally by its ohmic losses.\\

Finally, the last figure summarizes the different phenomena involved in the coupling of the metal particle with the photonic crystal for a radius varying between 0 and 150 nm by step of 10 nm, with few representative magnetic field distributions. Transmission and absorption spectra have been plotted side-by-side in log scale. As previously explained, for small radius, the ring mode is not strongly affected by the particle and shifts to the blue ($R$ branch), whereas the PC-plasmon modes (1) and (2) appears for radius larger than 110 nm. In between, we can see one mode which is linearly red-shifted with $R$: it corresponds to the octupolar, $m=4$, mode of the nanowire in the homogeneous matrix. The hexapolar $m=3$ mode appears at $\lambda=680$ nm for $R=60$ nm, but shifts beyond 700 nm for larger radius. Actually, for small radius, these modes are not strongly affected by the photonic crystal as the plasmon mode does not strongly overlap with the air holes of the photonic crystal. A transition occurs at about $R=100$ nm where these modes start to combine with the photonic crystal to give the PC-plasmon modes (1) and (2) previously described, where they show much narrower than modes $m=3$ and $m=4$.
\begin{figure}[h]
    \includegraphics[width=8.5cm]{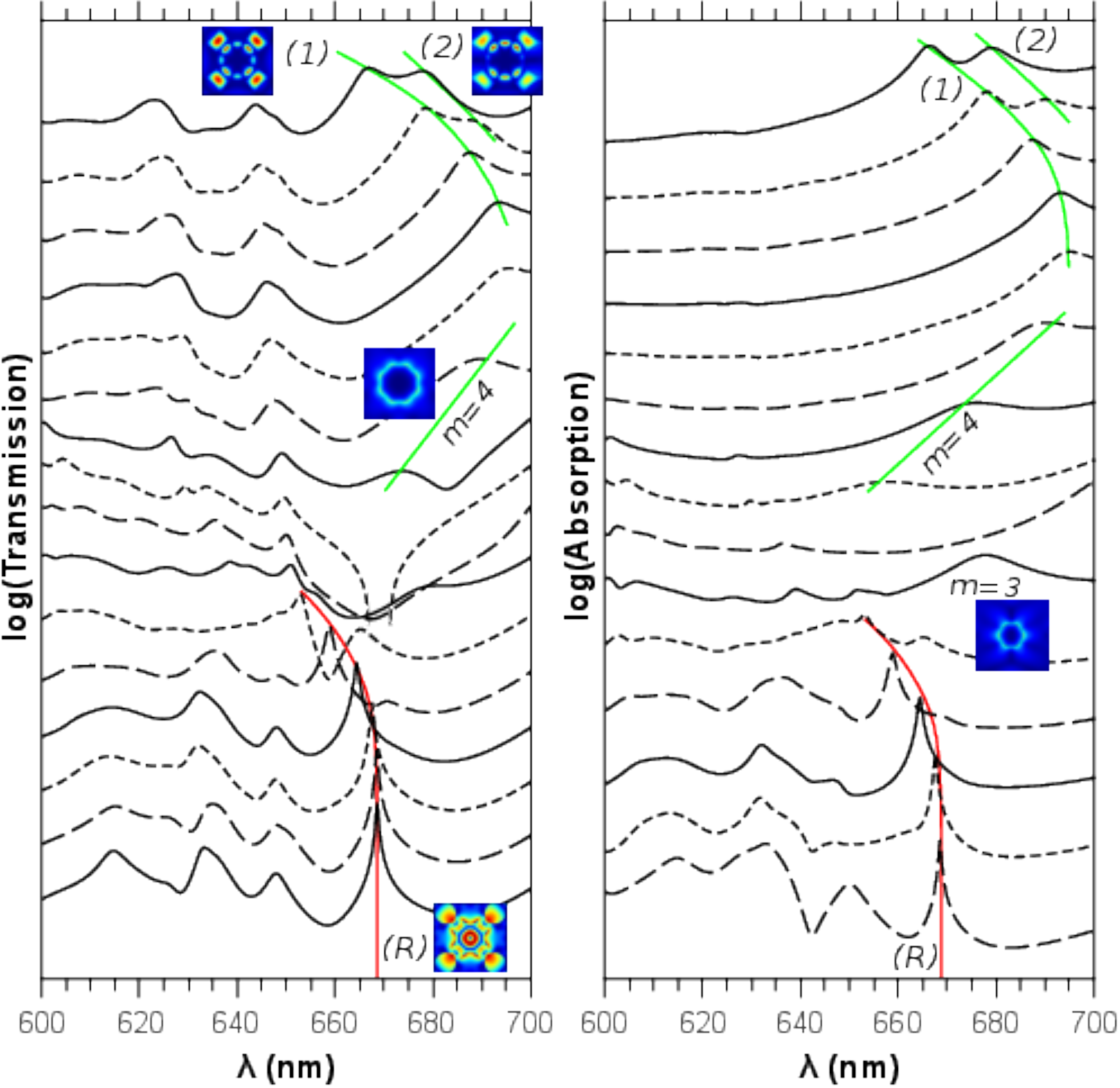}
    \caption{Evolution of absorption in the metal nanowire, as a function of the wavelength and the particle radius.}\label{fig:fig8}
\end{figure}

\section{Summary}
In this work, we have investigated the optical properties of a 2D metal nanowire placed at the center of a localized cavity in a 2D photonic crystal. The nature of the modes, their wavelength, together with their evolution with the particle radius is very dependent on how the diameter compares to the grating period. The ring photonic mode of the PC cavity has his shape essentially unaffected by a particle of few ten nanometers diameter, while its wavelength blue-shifts. However, for a particle with diameter about the lattice constant, the coupled system present very narrow resonances corresponding to compound PC-plasmon modes, where the narrowing of the resonance is attributed to the decay of the radiative losses of the plasmon particle by the photonic crystal. These results open interesting applications in areas where tailoring the plasmon response of nanoparticle is required, as in LSP resonance biosensing or high-speed photonic devices. Beside, such a system is a choice platform for investigating the coupling of localized plasmon modes with elastic waves, as it has previously been demonstrated that PC can leads to enhanced interaction between phononic and photonic waves \cite{Rolland2012,El-Jallal2014}. Similarly, such a compound photonic crystal/metal particle system could show enhanced interaction between elastic and plasmonic waves \cite{Large2009,Kelf2013,Tripathy2011}, allowing to achieve mechanical control of the plasmon response at a sub-micrometer scale.

\end{document}